\documentclass[submission,copyright,creativecommons]{eptcs}
\providecommand{\event}{TFP 2018}

\usepackage{listings}
\usepackage{pgfplots}
\usepackage{url}
\usepackage{xcolor}
\usepackage{fontawesome}
\usepackage{relsize}
\usepackage{pgfplotstable,filecontents}
\usepackage{float}
\usepackage{underscore}

\usepackage[section]{placeins}

\pgfkeys{
  /pgf/number format/read comma as period=true
}

\def\titlerunning{Investigating Compilation Errors of Students Learning Haskell}
\def\authorrunning{B. N\'emeth, E. Choi, E. Makihara, H. Iida}

\usetikzlibrary{patterns}
\usetikzlibrary{arrows}
\usepgfplotslibrary{statistics}

\providecommand{\urlalt}[2]{\href{#1}{#2}}
\providecommand{\doi}[1]{doi:\urlalt{http://dx.doi.org/#1}{#1}}

\begin{document}

\title{Investigating Compilation Errors \\of Students Learning Haskell\thanks{This work was partially supported by a mobility scholarship of the first author at Nara Institute of Science and Technology in the frame of the Erasmus Mundus Action 2 Project TEAM Technologies for Information and Communication Technologies, funded by the European Commission. This publication reflects the view only of the authors, and the Commission cannot be held responsible for any use which may be made of the information contained therein.}}

\author{Boldizs\'ar N\'emeth \institute{E\"otv\"os Lor\'and University} \email{nboldi@elte.hu} \and
Eunjong Choi \qquad\qquad\qquad Erina Makihara \qquad\qquad\qquad Hajimu Iida
\email{choi@is.naist.jp \qquad\qquad makihara.erina.lx0@is.naist.jp \qquad\qquad iida@itc.naist.jp}
\institute{Nara Institute of Science and Technology}}


\maketitle

\begin{abstract}
While functional programming is an efficient way to express complex software, functional programming languages have a steep learning curve. Haskell can be challenging to learn for students who were only introduced to imperative programming. It is important to look for methods and tools that may reduce the difficulty of learning functional programming. Finding methods to help students requires understanding the errors that students make while learning Haskell.

There are several previous studies revealing data about Haskell compiler errors, but they do not focus on the analysis of the compiler errors or they only study a certain kind of compiler errors.

This study investigates compilation errors of novice Haskell students and make suggestions on how their learning efficiency can be improved. Unlike previous studies we focus on uncovering the root problems with the student solutions by analysing samples of their submissions.
\end{abstract}

\section{Introduction}

The Haskell programming language \cite{haskell} is an advanced, purely functional programming language with static type system and lazy evaluation. Thanks to Haskell's expressive type system, many programming errors are caught by the compiler. It has been reported that programming languages with strong type systems are more resilient to bugs \cite{programming-languages-quality}.

Regardless of the benefits of using Haskell, it is considered to be a difficult programming language to learn. The causes of this belief may be related to its strict type checking, functional nature (that is difficult to understand for students who have been only introduced to imperative programming languages), and non-intuitive lazy evaluation.

Our motivation is to help students at overcoming the challenge of understanding and solving compiler errors. To do this, we must understand what kind of errors they make. This study investigates programming mistakes that results in a compiler error of the first year of bachelor students at E\"otv\"os Lor\'and University during a Haskell course.

Since compiler errors that arise in the first steps of the compilation process, namely lexical and syntactical errors, are easier to detect, quantitative data on these errors are more thoroughly covered \cite{baby-talk}. Our research focuses on complex errors found in later stages of the compilation. We focus on the errors that are the most frequently encountered by students, to provide usable solutions to their problems.

The goal of this study is to analyse the programming mistakes that students make while learning Haskell. In order to tackle the question of why Haskell is hard to learn we have to focus on the obstacles the students encounter while learning the language, and how they influence the errors the students make. Since learning a programming language is only achievable through practice, it is best to inspect the results of the students in a practical exercise, where they have to put their knowledge of the language into actual source code.

We are interested in providing results on how students can be helped during their study of functional programming. The information that we find in the students' compilation errors can be applied in many ways.

\begin{itemize}
\item The analysed data is valuable feedback for the educators. The results of our investigation could be applied to other courses with similar curriculum. By identifying the topics that are the hardest for the students it is possible to spend more time on these challenging topics, or postpone them until the students are more familiar with the basic language concepts.
\item Our research can aid compiler developers in improving the error messages of the compiler. By describing which errors are most frequently encountered by students and which of them are misleading or confusing for them we can point out error messages that could be improved.
\item Some of the students problems can be overcome with more sophisticated tooling. Our research can show which problems need the most attention and can guide further research on automated tools that, for example, describe the compiler errors to the students or make suggestions on how to correct them.
\end{itemize}

\section{Related work}

The process of learning functional programming has already been examined by several researchers.

Singer and Archibald collected data from an online course on Haskell and identified some common syntactic error patterns \cite{baby-talk}. However, the data presented in the paper only describes the frequencies of lexical and syntactic errors. The errors that are encountered in later stages of the compilation are not analysed. They mention their intent to perform a type-level analysis on the data, but as of today they have not published such results. They conclude that richer programming tools would be beneficial for their students.

Heeren et al. present an alternative Haskell compiler called Helium \cite{helium}, that is specifically targeted for students learning the language. It supports a subset of the Haskell language and provides better error messages than other Haskell compilers and automatic corrections to the students. Their paper contains a breakdown of error types over the course of seven weeks. Their approach of categorization of errors is similar to ours, their results suggest that the main cause of compiler errors is type related errors. However they offer no further breakdown of the error categories.

Gerdes, et al. describe an interactive tutor for Haskell \cite{ask-elle}. It supports interactive development with holes and gives specific advice by matching the student solution to the annotated examples. Guiding the student to solve the exercise step-by-step has an advantage over simply giving them tasks to do.

Pettit and Gee report in their paper that by simply making more useful and customized error messages for certain common programmer mistakes will not make students less likely to repeat the same errors \cite{enhanced-compiler-errors}, but their results are not conclusive. In a recent paper on static analysis tools, Barik argues that these tools should explain the steps they performed \cite{explain-anomalies}. A study with eye tracking supports that understanding error messages is vital for the success of a programming task and as difficult as reading source code \cite{read-compiler-errors}.

\section{Analysis method}

\subsection{The analysed dataset}

We investigated student errors in a one-semester course on functional programming. More than 120 first year undergraduate students of E\"otv\"os Lor\'and University participated in the course. They only have basic imperative programming experience in a statically typed programming language. The functional programming course consists of 12 lectures and practice sessions where the students use Haskell to solve exercises for about 60-70 minutes at a time, with homework and assignments between classes. Exercises for the practice sessions are short, typically requiring the student to write the implementation of one function (1-5 lines) given the type signature. For the practice sessions the students are divided into groups of 20. Among the 12 weeks of the course, we analyzed nine weeks (from second to tenth) where the attendance was high enough to offer a representative sample.

The curriculum of the course covers the basics of the Haskell language. The practice sessions start with integer and boolean arithmetic. After that, lists are introduced including the list comprehension syntax. This is followed by function definitions, pattern matching with a focus on recursion and using pattern guards. At the end of the semester higher-order functions are introduced, concentrating on list-processing functions, like folding. The precise scheduling of these topics varied between different groups.

The dataset used in this study is recorded on an interactive website (English version accessible \cite{lambdapage}) the students use for solving exercises and assignments. Not only the finished exercises are recorded but also every step the student took toward that final submission.

To make it easier to start the exercises, the students submitted code fragments instead of complete Haskell modules. The rest of the module was generated for them, including type signatures for the functions they had to write. This is somewhat limiting, for example, the students could not add a new import declaration to the module, but we provided them with a modified version of the standard Haskell Prelude (the automatically accessible parts of the standard library), and additional modules where the exercises required including them.

The website does not require accounts to be created to use it. Since students are using different computers each time they use the website, the lack of accounts makes it impossible to track the performance of individual students over the semester. It is, fortunately, not in the scope of our research. By taking advantage of how the site keeps student sessions, the interactions of a single student with the system can be recollected for the duration of a single class. The timestamps of the interactions are recorded, we are analysing how long it takes for the students to fix certain errors. We are using the timestamped sequences of interactions to analyse the students reactions to error messages. The dataset is anonymous, no personal information is present.

The infrastructure used for teaching Haskell includes an automated testing environment. The student submissions were compiled using the Glasgow Haskell Compiler \cite{ghc} (GHC), version 8.0.2. The system knows which exercises do the submissions associated with. Not all exercises are tested but compilation errors are reported for each one. The automated responses about the compilation errors and test results of the server are also recorded.

\begin{figure}[ht]
\centering
\caption{Collection of data for the study}
\begin{tikzpicture}[ line/.style={thick,->,>=stealth',font=\itshape\small},
                   node distance=3cm,mynode/.style={font=\relsize{4}}]
    
    \node[mynode,label={[font=\bfseries]left:Student}](student){\relsize{6}\faChild};
    \node[mynode,label={[font=\bfseries]above:Site},above right of=student](site){\faFirefox};
    \node[mynode,label={[font=\bfseries]right:Server},below right of=site](ghc){\faGears};
    \node[mynode,label={[font=\bfseries]below:BEAD},below right of=student](bead){\faSave};
    \node[mynode,right of=bead,label={[font=\bfseries]below:Teacher}](teacher){\faCoffee};
    \node[mynode,right of=site,label={[font=\bfseries]above:Logs}](logs){\faDatabase};
    \node[mynode,label={[font=\bfseries]right:Results},right of=ghc](results){\faFileText};

    \path[line] (student) edge[bend left] node[left] {submission} (site);
    \path[line] (student) edge node[left,align=left] {final\\submission} (bead);
    \path[line] (site) edge[bend left] node[right] {results} (student);
    \path[line,<->] (site) edge node[right,align=right] {compile\\\& run} (ghc);
    \path[line,<->] (bead) edge node[right,align=left] {\hspace{6pt}compile\\\& run} (ghc);
    \path[line] (bead) edge node[below] {results} (teacher);
    \path[line] (site) edge node[above] {logging} (logs);
    \path[line] (logs) edge node[right] {analysis} (results);
\end{tikzpicture}
\label{fig:data-flow}
\end{figure}
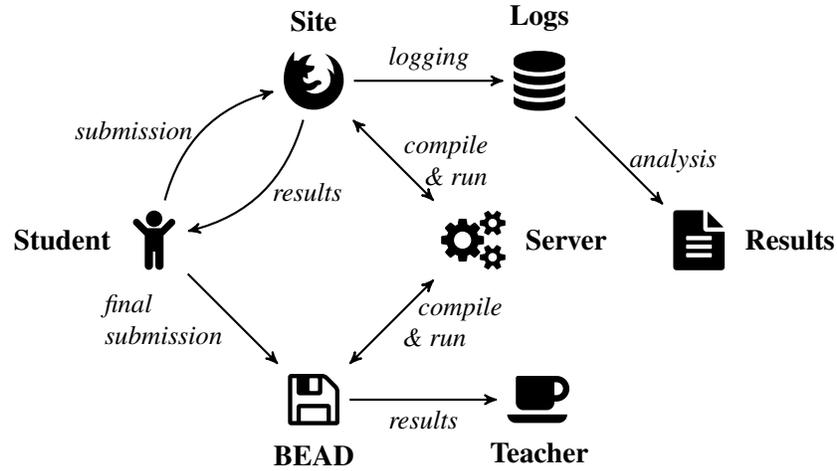

Figure \ref{fig:data-flow} presents the setting in which the students tried and submitted their solutions. They are using the interactive website to develop their solutions to the exercises. Their submissions are compiled and run, comparing the output to the expected results. Each compilation is a separate solution, so one student can easily generate 30-60 solutions during a single session. The log files generated by the interactive website are analysed in this paper. The students submit their final submissions as homeworks and assignments to a different system (BEAD), but that is not part of our dataset. The site logs both the student submissions and the responses from the server (the results of compiling and running the solution).

\subsection{Steps of the investigation}

To summarize our analysis of the extracted dataset, we present it in four steps:

\begin{enumerate}
\item We analyse how the success ratio of the student submissions changes in the course of the semester.
\item We assign compiler errors into categories based on the error messages and check how the ratio of errors in these categories change over the semester.
\item By analysing the sequences of submissions from students we measure how long it takes for the students to correct different kind of compiler errors.
\item We sample the whole dataset to determine the root causes of the most common higher-level errors.
\end{enumerate}

In some cases GHC error messages do not reveal the actual cause of the error. We also sampled the compiler errors to detect and categorize the root causes. We noticed that the cause of some of the error messages from later stages of compilations originate from syntactical mistakes on the students part. We tracked how the frequency of the syntactically caused problems change over the semester.

As a comparison to our dataset, we run the analysis on the dataset used by Singer and Archibald in their paper Functional Baby Talk (\emph{FBT} dataset in the following) \cite{baby-talk}. Their dataset is collected during an online course, so it complements our dataset that is collected during a traditional university class.

\section{Result of the investigation}

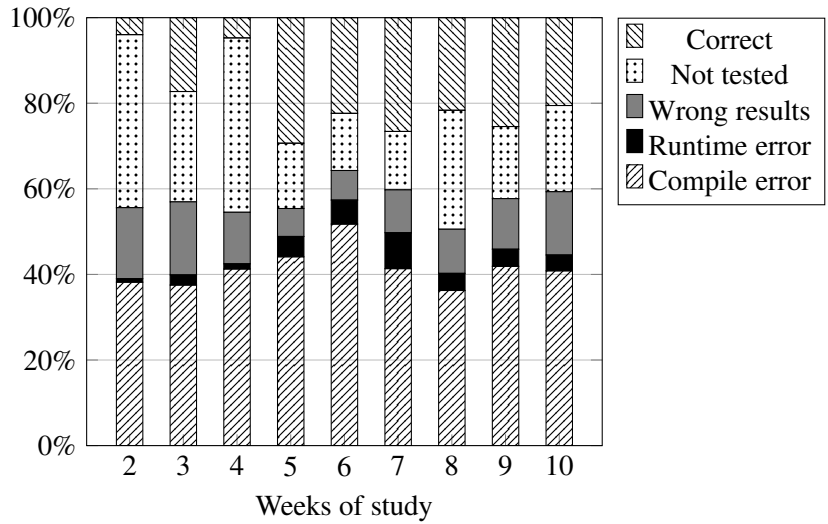
\begin{figure}[ht]
\centering
\caption{Results of student solutions}
\begin{tikzpicture}
\begin{axis}[
	xtick=data,
	ybar stacked,
	legend pos=outer north east,
	reverse legend=true,
	xlabel=Weeks of study,
	ymajorgrids,
    ymin=0,
    ymax=1,
    yticklabel={\pgfmathparse{\tick*100}\pgfmathprintnumber{\pgfmathresult}\%},
]
\addplot[pattern=north east lines] table [x=week, y=compile-error, col sep=semicolon] {success.csv};
\addplot[fill=black] table [x=week, y=runtime-error, col sep=semicolon] {success.csv};
\addplot[fill=gray] table [x=week, y=wrong-result, col sep=semicolon] {success.csv};
\addplot[pattern=dots] table [x=week, y=not-a-test, col sep=semicolon] {success.csv};
\addplot[pattern=north west lines] table [x=week, y=correct, col sep=semicolon] {success.csv};

\legend{Compile error,Runtime error,Wrong results,Not tested,Correct}
\end{axis}
\end{tikzpicture}
\label{fig:success}
\end{figure}

\begin{table}[ht]
\centering
\small
\caption{Results of student solutions}
\vspace{5pt}
\begin{filecontents*}{success-nums.csv}
week;compile-error;runtime-error;non-termination;wrong-result;correct;not-a-test
2;2195;36;11;954;228;2324
3;1513;81;15;688;696;1040
4;1423;36;6;415;163;1405
5;743;79;1;110;494;257
6;1116;121;0;149;482;288
7;930;188;1;225;598;307
8;1610;178;0;458;961;1235
9;2801;264;4;790;1701;1127
10;4534;396;15;1641;2279;2236
\end{filecontents*}
\pgfplotstabletypeset[col sep=semicolon,
                      before row=\hline,
                      every last row/.style={after row=\hline},
                      column type={|p{1.3cm}},
                      columns/week/.style={column name=Week},
                      columns/compile-error/.style={column name=Compile errors},
                      columns/runtime-error/.style={column name=Runtime error},
                      columns/non-termination/.style={column name=Non termination},
                      columns/wrong-result/.style={column name=Wrong results},
                      columns/correct/.style={column name=Correct},
                      columns/not-a-test/.style={column type={|p{1.3cm}|},column name=Not a test},
                     ]{success-nums.csv}
\label{tab:success}
\end{table}

Table \ref{tab:success} and Figure \ref{fig:success} shows the frequency of student submissions resulting in compiler errors, runtime errors, as well as wrong results and correct results. (``Not tested'' means that the solution compiled, but there is no test associated with the exercise. ``Correct'' means that the solution passed the tests of the exercise.) The results were obtained by mechanically classifying the compiler output. On Table \ref{tab:success} we can see that the number of submissions peaked in the last week of the semester. However Figure \ref{fig:success} shows, that the ratio of successful submissions does not really change during the course of the semester. This implies that the difficulty of the exercises is balanced by the students growing proficiency in Haskell.

In the FBT dataset the rate of type-correct submissions is also constant during the course, but it is much higher (70\% - 80\% against our 50\% - 60\%). We assume that this is caused by the different difficulties and structures of the exercises. The activity is declining during the course peaking on the second week and steadily decreasing to about 5\% of that peak by the last week in the FBT dataset. Compared to this, in our dataset the activity peaks in the last weeks when students are learning for the exams.

\begin{table}[ht]
\centering
\small
\caption{Categories of students errors based on the error messages}
\vspace{5pt}
\begin{filecontents*}{categories-nums.csv}
week;parse;rename;type
1;950;448;971
2;507;458;694
3;627;324;536
4;228;289;301
5;254;211;738
6;219;192;565
7;365;382;980
8;594;631;1729
9;982;1045;2851
\end{filecontents*}
\pgfplotstabletypeset[col sep=semicolon,
                      before row=\hline,
                      every last row/.style={after row=\hline},
                      column type={|p{1.3cm}},
                      columns/week/.style={column name=Week},
                      columns/parse/.style={column name=Parse},
                      columns/rename/.style={column name=Naming},
                      columns/type/.style={column type={|p{1.3cm}|},column name=Type check},
                     ]{categories-nums.csv}
\label{tab:error-cats}
\end{table}

Based on the phase of compilation when they are reported, we grouped compiler errors into three distinct categories: \emph{syntactic} (lexical and syntax-related) errors are found in the first stage of compilation, while \emph{name-related} errors (missing names, name collisions) and \emph{type-check} errors (errors found during type checking) found in the second main stage of compilation. Please note that errors in earlier stages can hide errors in later stages. The frequencies of these error categories can be seen in Table \ref{tab:error-cats}.

However, we found that not all compiler errors are correctly categorized by their error messages. Because of Haskell's simple and flexible syntax, some syntactic errors that the student made accidentally resulted in syntactically correct programs. In these cases the mistake produced an error in a later stage of the compilation, producing a misleading error message. We call these errors \emph{false semantic errors}.

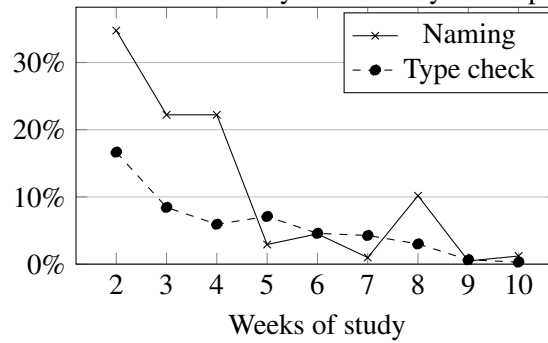
\begin{figure}[ht]
\centering
\caption{Student errors falsely related to syntactic problems}
\begin{tikzpicture}
\begin{axis}[
    yticklabel={\pgfmathparse{\tick*100}\pgfmathprintnumber{\pgfmathresult}\%},
    xlabel=Weeks of study,
    ymin=0,
    ymajorgrids,
    xtick=data,
    height=5cm,
    width=8cm,
]
\addplot[solid,mark=x] table [x=week, y=rename, col sep=semicolon] {miscategorized.csv};
\addplot[dashed,mark=*] table [x=week, y=type, col sep=semicolon] {miscategorized.csv};

\legend{Naming,Type check}
\end{axis}
\end{tikzpicture}
\label{fig:error-syntactic}
\end{figure}

\begin{table}[ht]
\centering
\small
\caption{Student errors falsely related to syntactic problems}
\vspace{5pt}
\begin{filecontents*}{miscategorized-nums.csv}
week;typecheck;typecheck-syntax;rename;rename-syntax
2;258;43;164;57
3;225;19;180;40
4;202;12;135;30
5;141;10;137;4
6;240;11;111;5
7;211;9;102;1
8;266;8;167;17
9;308;2;209;1
10;332;1;248;3
\end{filecontents*}
\pgfplotstabletypeset[col sep=semicolon,
                      before row=\hline,
                      every last row/.style={after row=\hline},
                      column type={|p{1.8cm}},
                      columns/week/.style={column name=Week},
                      columns/typecheck/.style={column name=Type-check errors},
                      columns/typecheck-syntax/.style={column name=False Type-check errors},
                      columns/rename/.style={column name=Naming errors},
                      columns/rename-syntax/.style={column type={|p{1.8cm}|},column name=False Naming errors},
                     ]{miscategorized-nums.csv}
\label{tab:error-syntactic}
\end{table}

To estimate the frequency of false semantic errors, we manually sampled the errors from the later stages of the compilation, and grouped them into real semantic errors and syntax-related, false semantic errors. The results can be seen on Figure \ref{fig:error-syntactic} and in Table \ref{tab:error-syntactic}. It is important to note that Table \ref{tab:error-cats} contains the errors based on their error message only, the false semantic errors are not removed from the data presented in it. It is also important that Table \ref{tab:success} summarizes submissions while Table \ref{tab:error-cats} and Table \ref{tab:error-syntactic} detail the number of errors, since more than one error can appear in a single submission.

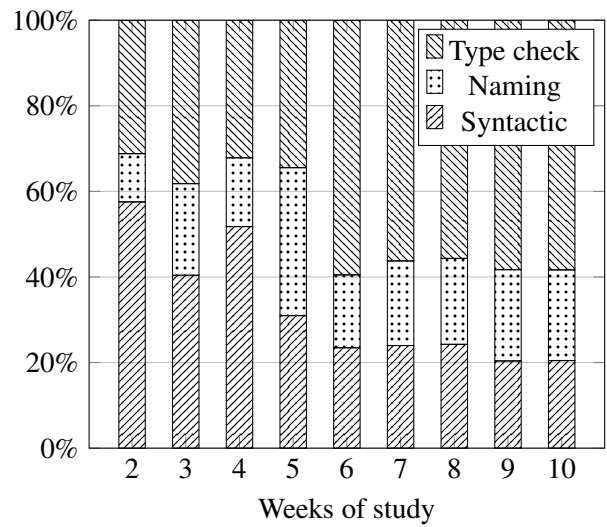
\begin{figure}[ht]
\centering
\caption{Categories of students errors based on the error messages}
\begin{tikzpicture}
\begin{axis}[
    ymin=0,
    ymax=1,
    yticklabel={\pgfmathparse{\tick*100}\pgfmathprintnumber{\pgfmathresult}\%},
	ymajorgrids,
	xtick=data,
	xlabel=Weeks of study,
	reverse legend=true,
	ybar stacked
]
\addplot[pattern=north east lines] table [x=week, y=parse, col sep=semicolon] {categories-corrected.csv};
\addplot[pattern=dots] table [x=week, y=rename, col sep=semicolon] {categories-corrected.csv};
\addplot[pattern=north west lines] table [x=week, y=type, col sep=semicolon] {categories-corrected.csv};

\legend{Syntactic,Naming,Type check}
\end{axis}
\end{tikzpicture}
\label{fig:error-cats}
\end{figure}

Figure \ref{fig:error-cats} shows the frequency of the compilation errors in different stages of compilation. The results were obtained by mechanically classifying the compiler output, but false semantic errors are manually reclassified in the result. As it can be seen on this figure, name and type related errors are responsible for a large part of the compilation errors. Moreover, during the semester type-related errors become even more frequent. By week 5, more than half of the error messages are produced by type errors. This corresponds to the fact that in the second part of the semester more complex types (higher-order functions, polymorphism) are introduced.

The FBT dataset provided results similar to ours, but without syntactic errors being dominant in the first part of the course. We assume that this is because the course was shorter, and the used Haskell syntax was more restricted, than our dataset. Interestingly in the FBT dataset, name-related errors are decreasing and syntax errors are increasing. However low activity in the last week may be distorting the results. Similar results are presented by Heeren et al. using Helium \cite{helium}, with type errors causing more than 50\% of all errors.

Figure \ref{fig:solve-times} shows how long it takes for the students to correct one type of compilation errors as a box plot. We measured the time difference between each given error and the submission which passes that compilation stage. The measurement is done automatically from the recorded logs and the figure shows the aggregated results for the whole semester. For syntax errors we measured the time until all syntax errors disappear, for name and type related errors, we wait until the first submission that does not result in a compilation error. We chose not to focus on the elimination of individual errors, since a wrong solution to a given error may result in a similar, but somewhat different error. We, however, took into account all submissions in a session that result in a compilation error, not just the first one. As a comparison we included the correction times of runtime errors as well (including runtime exceptions, non-termination and incorrect results). Only those interactions are incorporated where the error was eventually fixed.

\begin{figure}[ht]
\centering
\caption{Correction times of errors in different categories}
\begin{tikzpicture}
\begin{axis} [
  ytick={1,2,3,4},
  yticklabels={Syntactic, Naming, Type check, Runtime},
  xlabel=Correction time (seconds), 
  height=5cm,
  xmin=0,
  xmajorgrids,
  xticklabel={\pgfmathprintnumber[assume math mode=true]{\tick}s}, 
  width=10cm]

\addplot+ [black, thick, boxplot prepared={lower whisker=1, lower quartile=6, median=14, upper quartile=46, upper whisker=199}] coordinates {};
\addplot+ [black, thick, boxplot prepared={lower whisker=2, lower quartile=8, median=22, upper quartile=68, upper whisker=262.75}] coordinates {};
\addplot+ [black, thick, boxplot prepared={lower whisker=3, lower quartile=11, median=32, upper quartile=88, upper whisker=282}] coordinates {};
\addplot+ [black, thick, boxplot prepared={lower whisker=4, lower quartile=17, median=46, upper quartile=123, upper whisker=331.45}] coordinates {};
\end{axis}

\end{tikzpicture}
\label{fig:solve-times}
\end{figure}

The first author, as a Haskell instructor, experienced that complex error messages often get the students to stop progressing for an unexpected amount of time. As the students are trying to understand the error message, they are not finding the often easy-to-spot problems in their submission. Usually these problems are solved by the instructor who is asked by the student, or notices that the student's progress is halted by the problem.

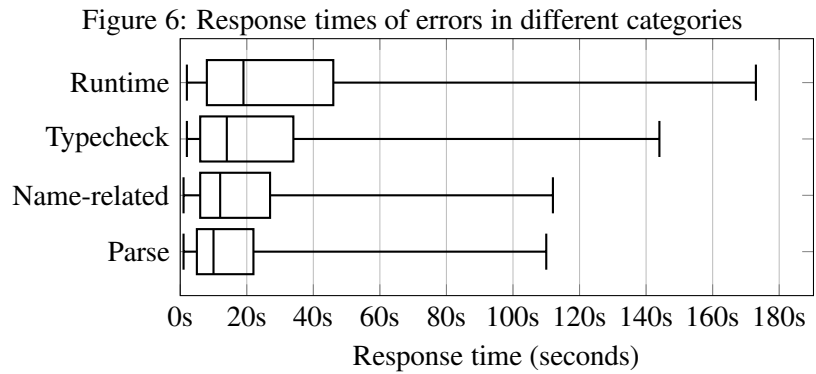
\begin{figure}[ht]
\centering
\caption{Response times of errors in different categories}
\begin{tikzpicture}
\begin{axis} [
  ytick={1,2,3,4},
  yticklabels={Parse, Name-related, Typecheck, Runtime},
  xlabel=Response time (seconds), 
  xmin=0,
  xmajorgrids,
  xticklabel={\pgfmathprintnumber[assume math mode=true]{\tick}s},
  height=5cm, 
  width=10cm]

\addplot+ [black, thick, boxplot prepared={lower whisker=1, lower quartile=5, median=10, upper quartile=22, upper whisker=110}] coordinates {};
\addplot+ [black, thick, boxplot prepared={lower whisker=1, lower quartile=6, median=12, upper quartile=27, upper whisker=112}] coordinates {};
\addplot+ [black, thick, boxplot prepared={lower whisker=2, lower quartile=6, median=14, upper quartile=34, upper whisker=144}] coordinates {};
\addplot+ [black, thick, boxplot prepared={lower whisker=2, lower quartile=8, median=19, upper quartile=46, upper whisker=173}] coordinates {};
\end{axis}

\end{tikzpicture}
\label{fig:response-times}
\end{figure}

In this study we quantified this phenomenon by measuring how long does it take for the student to come up with the next submission after an error. As it can be seen on Figure \ref{fig:response-times}, the result is more striking then in the case of solution times. The time it takes for students to come up with an answer to a type check error close to the time it takes to create a new solution to a runtime problem. It is not unreasonable to say that understanding the type-check error messages is as hard for the students as mentally debugging the runtime behavior of their solution.

\subsection{Inspecting the root causes of semantic errors}

Since Figure \ref{fig:error-cats} confirms that after learning the basic syntax, the students make compilation errors that are found in later stages of the compilation, we focus on these errors. We saw on Figure \ref{fig:error-syntactic} that the error messages can be misleading in some cases.

To inspect the nature of these errors further, hereafter, we show the five most frequent root causes for name-related and typecheck errors. The results are made by sampling the compile errors of the given category. The root causes of errors are determined manually. They are relevant with 5\% confidence interval and a .95 confidence level.

\subsubsection{Name-related errors}

\paragraph{Global name mistake (\textbf{23.8\%} of name-related errors)}

Since students are not familiar with the Haskell environment, name-related errors are often caused by mistakes in referencing entities of the standard library or the testing environment. 

The following is an example of a mistake in referencing an imported definition. The student has written \lstinline{True} and \lstinline{False} in lowercase.

\begin{lstlisting}
isSingleton [a] = (-\framebox{true}-)
isSingleton _ = (-\framebox{false}-)
\end{lstlisting}

\paragraph{Definition name mistake (\textbf{14.2\%} of name-related errors)}

In other cases the problem is not accessing global entities but the correct spelling of the name of the function that is defined. In the exercises, the type signature is often already given. In this cases when the type signature and the name of the binding does not match, the compiler reports that it cannot find the binding that corresponds to the type signature present.

As an example of this error, in this submission there is a simple typing mistake in the name of the defined function. The type signature that is generated as part of the exercise is \lstinline{isCircled :: Cell -> Bool}.

\begin{lstlisting}
(-\framebox{isCricled}-) x = label x == Circled
\end{lstlisting}

\paragraph{A syntactic mistake (\textbf{12.5\%} of name-related errors)}

As it was shown earlier errors from later stages of compilation result from simple syntactical mistakes on the students part. For example the list expression \lstinline{[10,9..-10]} is not correct, because the operator \lstinline{..} and the prefix negative sign are not separated, the compiler actually looks for an operator named \lstinline{..-}, which does not exist. Other common causes of syntactical errors are discussed in Section \ref{discussion}.

\paragraph{Local name mistake (\textbf{6.7\%} of name-related errors)}

Referencing bindings and variables defined by the student are less likely to be mistaken than references to parts of the environment. This is not surprising, since students know the source code they have written better than the standard library.

The following example presents an error in referencing a local name defined by the student. The name \lstinline{n} is bound for the parameter of the function, but the student uses the name \lstinline{e} as a mistake.

\begin{lstlisting}
elem n [] = False
elem n (x:xs)
 | (-\framebox{e}-) == x = True
 | otherwise = elem n xs
\end{lstlisting}

Probably most of these mistakes are just the result of forgetfulness and quickly corrected.

\paragraph{Misunderstanding list comprehensions (\textbf{5.8\%} of name-related errors)}

The students seem to have a lot of problems with the correct syntax of list comprehensions. These syntactic structures are not present in most of the conventional programming languages, but taught early in the class. This might be the reason why students are struggling with this language element.

In some cases it is clear that the student does not understand the structure of this language element as follows.

\begin{lstlisting}
sumSquaresTo n = sum[i|i<-[1..](-\framebox{*i}-)]
\end{lstlisting}

In other cases, like in the following example, the cause of the problem might have been just a simple typing error:

\begin{lstlisting}
[n|m<-[1..],n(-\framebox{<}-)[1..m]]
\end{lstlisting}

\subsubsection{Type-related errors}

For type-related errors, the distinction between root causes is harder than name-related errors, since the problems are more complex. In some cases it is only possible to determine what kind of modifications would be necessary to compile the student submission.

\paragraph{List type mismatch (\textbf{20\%} of type-related errors)}

A very common mistake for students is to confuse list and scalar types. In other cases lists of different dimensions are the cause of the problem. The root error is often an application of an incorrect operation that results in hard-to read type errors.

The following describes an example of this kind of problem. The task was to define the \lstinline{cutRepeated} function that returns the longest prefix of the list that has no adjacent identical elements. However, the student converted the expression \lstinline{a:(cutRepeated_ x xs)} (of type \lstinline{[a]}) into a one-element list. The resulting expression has the type \lstinline{[[a]]} that conflicts with the type of the expression \lstinline{[a]}.

\begin{lstlisting}
cutRepeated_ a (x:xs)
    | a == x = [a]
    | otherwise = (-\framebox{[}-)a:(cutRepeated_ x xs)(-\framebox{]}-)
\end{lstlisting}

The resulting type error is a bit complicated, and does not describe the problem well: 

\begin{lstlisting}
Cannot construct the infinite type: t ~ [t].
In the expression: a : (cutRepeated_ x xs).
\end{lstlisting}

\paragraph{Function argument missing (\textbf{8.6\%} of type-related errors)}

The second most frequent type-related mistake of students is to forget to supply one of the arguments for a function application. Take the example of the following submission that is trying to implement the \lstinline{count} operation using the \lstinline{filter} operation. 

\begin{lstlisting}
count f l = length ((-\framebox{filter l}-))
\end{lstlisting}

The student forgot to pass the argument \lstinline{f} to the \lstinline{filter} function. The error message of the compiler tells the user that the argument of the length function should be a list, but since the function application is not complete, it is a function instead. There is also an additional error message because the first argument of the function is missing and the parameter \lstinline{l} is in the wrong place.

\paragraph{Simple type mismatch (\textbf{8.6\%} of type-related errors)}

Confusion between simple types, like \lstinline{Integer}, \lstinline{Int}, \lstinline{Double}, \lstinline{Char}, \lstinline{String} or \lstinline{Bool} are frequently encountered among the student submissions. This may be related to the fact that Haskell does not have automatic conversions between these data types, while implicit conversions are present in other programming languages.

A classic example of this type of error is a problem where the student tries to use the division \lstinline{/} operator on integrals, where it is only applicable to fractional numbers in Haskell.

\begin{lstlisting}
x ^ n
 | n==0      = 1
 | odd n     = x * x^(n-1)
 | otherwise = sqr x ^ ((-\framebox{n/2}-))
\end{lstlisting}

The student should use the \lstinline{div} operator instead that performs integer division. The exponentiation operators \lstinline{^}, \lstinline{^^} and \lstinline{**} are also causing frequent problems, because students tend to confuse them with each other.

\paragraph{Wrong operation applied (\textbf{8.4\%} of type-related errors)}

In many cases the students did not have knowledge of the available functions and did not find the correct one to apply. In the following example the student used the \lstinline{concat} function (that collapses lists of lists into one-dimensional lists), instead of the supposed \lstinline{++} operator that appends one list to the other.

\begin{lstlisting}
chunksOf _ [] = [] 
chunksOf n l@(x:xs) = [take n l] (-\framebox{`concat`}-) chunksOf n (drop n l)
\end{lstlisting}

The corrected version of this source code:

\begin{lstlisting}
chunksOf _ [] = [] 
chunksOf n l@(x:xs) = [take n l] ++ chunksOf n (drop n l)
\end{lstlisting}

\paragraph{Wrongly grouped parameters (\textbf{6.7\%} of type-related errors)}

Since in Haskell function application is written as the juxtaposition of the function and the arguments (\lstinline{f x} means apply the the function \lstinline{f} to the argument \lstinline{x}) it can be confusing for students who are accustomed to languages where function application is always marked by parentheses.

Take the function \lstinline{cutRepeated}, that returns the longest prefix of the list that has no adjecent identical elements:

\begin{lstlisting}
cutRepeated (x:y:xs)
	| x == y = [x]
	| otherwise = x: cutRepeated (-\framebox{y:xs}-)
\end{lstlisting}

The student probably meant to apply \lstinline{cutRepeated} to a list \lstinline{y:xs} (this explanation is supported by the lack of spaces in \lstinline{y:xs}), but since operators have lower precedence than function application, GHC interpreted the right-hand side as \lstinline{ x: (cutRepeated y) : xs }, which is the application of a list-processing function to a scalar value.

The solution is to simply wrap the expression \lstinline{y:xs} in parenthesis.

\begin{lstlisting}
cutRepeated (x:y:xs)
	| x == y = [x]
	| otherwise = x: cutRepeated (y:xs) 
\end{lstlisting}

\section{Discussions}
\label{discussion}

Based on the investigation results we can say that type errors are the most frequent errors after a few weeks of studying Haskell. While name-related errors mostly come from misspelled names, as it was reported by earlier studies \cite{helium}, type errors actually have many different causes. This could be the reason why the type errors are persistent in student submissions through the semester. They are also harder to correct than syntax or name-related errors.

We also found that errors caused by mistakes in list comprehensions are numerous. Since the list comprehension syntax is just an alternative to using list-processing functions, they could maybe rescheduled for a more advanced Haskell course. Another topic that caused a lot of problems for the students was using lists with various element types and dimensions.

From a compiler design viewpoint, it is problematic that some of the syntactic mistakes may cause problems in later compilation stages, with messages unrelated to the actual problem. Some of the instances of this problem often found in the student solution:

\begin{itemize}
\item Writing the negative sign not separated from other symbols, for example \lstinline{[0,-1..-10]}.
\item Confusing name and operator usage, trying to use operators in prefix form without parentheses \lstinline{foldl + 0 [1..10]}, or trying to use names as operators without backticks \lstinline{a mod b}.
\item Forgetting to put explicit multiplication operator in expressions: \lstinline{3x + 4y}.
\end{itemize}

We suggest targeting these problems with special cases of error reporting. This would involve changing the compiler to recognize these circumstances and respond with a more informative error message. The results of this study suggests that tools to help students identify the actual compilation problems in their code would be greatly beneficial for learning purposes.

\paragraph{Threats to validity}

The study was made using the data from one Haskell course of one university. It is certain that the curriculum followed affected the results of this study. Result from the FBT dataset were used to generalize the results. The system that provided our dataset was anonymous, so there is a some chance that people outside the course also used it and their submissions got into the dataset as well.

The categorization of the different causes of compile errors are based on the experience of the first author as Haskell programmer and instructor. The results may be affected by miscategorized errors.

\section{Conclusions}

In this paper, we studied the compilation errors made by students while using an interactive website during their beginner Haskell course. We analysed the data using several steps.

First, we looked at the overall rates of success and failure, and found that the ratio of compiler errors keeps the same during the semester.

We then divided compiler errors into three broad categories, syntactic, name-related and type check errors. The results show that at the start of the course, syntactic errors are the most frequent, but during the semester, type check errors become the major source of compiler errors. To make the results more precise than what is possible using only the error messages, we sampled the submissions to spot the errors introduced by miscategorized error messages.

We also reconstructed student sessions and used the data to measure how long it takes for the students to correct different errors. We used two metrics for this, the ``time to fix a problem category'' and the ``time to respond'' measurements. From the results it is clear that type errors are not just the most frequent of the different compiler errors, but they take the most time to fix.

Finally, we applied a sampling to the set of error messages to group them according to common root causes. We have shown the most common root causes for name-related and type check errors.

\paragraph{Future work} Following this research, our next goal is to design helping tools for the students. This would mean to use some external tooling to help students in solving compile errors or avoid them in the first place. We also look forward to analyse individual errors in more detail.

\end{document}